\documentstyle[twocolumn,aps,psfig]{revtex}
\tighten
\begin{document}

\title{
Competition of Superconductivity and Antiferromagnetism in a d-Wave
Vortex Lattice
}
\author{
Amit Ghosal, Catherine Kallin and A. John Berlinsky
}
\address{Department of Physics and Astronomy,
McMaster University, Hamilton, Ontario, L8S 4M1, Canada
}
\maketitle
\begin{abstract}
The d-wave vortex lattice state is studied within the framework of Bogoliubov-de
Gennes (BdG) mean field theory. We allow antiferromagnetic (AFM) order to
develop self-consistently along with d-wave singlet superconducting (dSC)
order in response to an external magnetic field that generates vortices.
The resulting AFM order has strong peaks at the vortex centers, and
changes sign, creating domain walls along lines where $\nabla \times j_s
\approx 0$. The length scale for decay of this AFM order is found to be much
larger than the bare d-wave coherence length, $\xi$. Coexistence of dSC and AFM
order in this system is shown to induce $\pi$-triplet superconducting order. 
Competition between different orders is found to suppress the local density of 
states at the vortex center and comparison to recent experimental findings is
discussed.
\end{abstract}

\date{\today}

\narrowtext
\section{Introduction}

Many of the anomalous properties of high $T_c$ superconductors (SC) 
are believed to be due to the competition between different kinds of
orders.  The electronic structure of vortices in the cuprates is one striking 
example of behavior which
is not understood and which may be influenced by competing or
coexisting order parameters.  
The electronic structure of s-wave vortices has been studied extensively by both
theoretical \cite{caroli,gygi} and experimental \cite{hess} methods which
established the existence of localized vortex states with very low energy.
The situation is more complex for the case of d-wave SC due to the presence
of nodes in the energy gap.  Earlier theoretical results of Wang and MacDonald
\cite{wang}, performed within the framework of Bogliubov-de Gennes (BdG)
calculations, predicted a strong and broad zero bias conductance peak (ZBCP)
in the local density of states (LDOS) at the vortex core. This ZBCP was
shown \cite{ichioka} to display a four-fold symmetric ``star-shape", oriented
towards the gap nodes and decaying as a power law with distance from the vortex
core. In contrast, experiments present a completely different picture.
Tunneling studies of both YBCO \cite{maggio} and BSCCO \cite{pan} show no
evidence of a ZBCP.  Furthermore, structure is observed in the low energy
differential conductance data which has been interpreted as the existence 
of low energy quasiparticle states at
$\pm 5.5$ meV (for YBCO) and at $\pm 7$ meV (for BSCCO).  These states persist
to long distances -- a size much larger than that of the vortex core.

There have been a number of different proposals for reconciling this
contradiction between theoretical predictions and experimental findings,
including effects of strong correlations and competing order parameters.
The role of strong correlations was emphasized by
Franz and Tesanovic \cite{franz} who argued that the correct description
of the normal state of high $T_c$ cuprates is a doped Mott insulator and
not a simple metal (as in BCS theory).  Hence, at the vortex core, where
superconductivity is suppressed, a gapped insulator should be expected.
In this case, the pseudogap behavior of the tunneling results at the core center
is a result of spin-charge separation. To address the correlation
effects on the structure of an isolated vortex, Han and Lee \cite{han} 
studied the $t-J$
model using resonant valence bond mean field theory. Their main result is the
prediction of two types of vortex core -- insulating or metallic, depending
on the model parameters. They also found evidence of staggered current
patterns \cite{han2} in the vortex core.  The pairing order parameter
near the vortex core was found to have $d_{x^2-y^2} + id_{xy}$ symmetry
\cite{han,franz2}.  A different argument for the absence of the ZBCP 
invokes the structure of the c-axis tunneling matrix element
\cite{wu}.  In a recent study, Berthod and Giovannini \cite{berthod} showed
that with increasing correlation strength the BCS core states become suppressed
and the spectra inside and outside the vortex core become similar.

An alternative scenario has been presented in which antiferromagnetic (AFM)
order is generated near the vortex core. 
Arovas and collaborators \cite{arovas} first studied this problem within the 
framework of SO(5)
theory and predicted that the vortex core becomes antiferromagnetic, and hence
insulating, at low temperature and relatively low doping.  Subsequently, a
BdG calculation of SO(5) vortices\cite{andersen} found
that the AFM order, although itself gapless within this model,  suppressed the ZBCP.
Recent works of
Sachdev and collaborators \cite{sachdev} argue that the system undergoes a
quantum phase transition from a pure d-wave superconducting (dSC) phase to a
mixture of dSC and spin density wave (SDW) phase as a function of the external
magnetic field. In this theory the dynamic spin fluctuations in dSC are
pinned by the vortex cores.  More recently Zhu and Ting \cite{ting1} have
studied a d-wave vortex lattice, allowing for AFM order, using BdG mean
field theory.  They find AFM order which alternates in sign from one
vortex to the next, and a splitting of the ZBCP due to the AFM order.

There have been a large number of recent experiments which suggest the emergence
of  AFM or SDW order on top of the dSC order in the cuprates in
an external magnetic field.  Using
inelastic neutron scattering, Lake {\it et al.} \cite{lake} found that
the magnetic field induces a large increase in the spectral density of low
energy spin fluctuations in LSCO near optimal doping. This demonstrates the
existence of a fluctuating magnetisation density with a periodicity of 8 lattice
spacings along the direction of the Cu-O bonds. Scanning tunneling microscopy (STM)
on slightly overdoped BSCCO \cite{hoffman} in the mixed state shows a
charge modulation in the LDOS data that has a periodicity of 4 lattice
spacings -- a result consistent with the neutron scattering study since the
periodicity of charge modulation is one half of that of the SDW modulation.
Field induced enhancement of the Bragg peak intensity has also been
observed by Khaykovich {\it et al.}\cite{khayk}, in an elastic neutron
scattering study of ${\rm La_2CuO_{4+y}}$, indicating a field induced static
AFM order.  High-field nuclear magnetic resonance (NMR) studies of YBCO
\cite{mitro} have found evidence for strong antiferromagnetic fluctuations 
inside the core.  Muon spin rotation measurements\cite{miller} have
also been interpreted as evidence of static magnetism
in the vortex core of Ortho-II ${\rm YBa_2Cu_3O_{6.5}}$. Understanding the
charge and spin modulation in the vortex lattice state is currently under
intensive theoretical investigation \cite{zhu,franz3,zhang2}.

Motivated by the discrepancy between theoretical predictions and experimental
observations, we study the structure and spectrum of a d-wave vortex lattice
using a BdG mean field theory which allows both static AFM order and dSC
order in a self-consistent manner.  
Our main results are: (1) In the lowest energy configuration, AFM order
has the same periodicity as the vortex lattice, in contrast to earlier
studies\cite{ting1}, although there are antiphase domain walls in the
AFM order within the unit cell; (2) triplet SC order is induced in the
presence of coexisting AFM and dSC, which approximately
satisfies the constraint predicted
by Zhang\cite{zhang}; (3) due to the presence of AFM order, the LDOS
exhibits structure which persists well beyond the vortex core size.

The organization of the paper is as follows. In Sec. II, we 
describe the class of models we study, and we also elaborate on the BdG method.
Sec. III discusses our results on the spatial structures of different order
parameters in the mixed state, and Sec. IV discusses the LDOS results.  Finally,
we present conclusions in Sec. IV.

\section{Model and Method}\label{sec:model}

We describe a 2D d-wave SC in an external magnetic field by the following
Hamiltonian :
\begin{equation}
{\cal H} = {\cal K} +
{\cal H}_{\rm int} + {\cal H}_{\mu},
\label {eq:ham}
\end{equation}
where the kinetic energy, 
\begin{equation}
{\cal K} = -t \sum_{<ij>,\alpha} \left(e^{i \phi_{ij}}
c_{i\alpha}^{\dag} c_{j\alpha} + h.c.\right),
\label {eq:KEham}
\end{equation}
 describes electrons, with spin
$\alpha$ at site $i$ created by $c_{i\alpha}^{\dag}$, hopping between
nearest-neighbors $\langle ij\rangle$ on a square lattice. 
The Peierl's phase factor
$\phi_{ij} = (\pi/\Phi_0) \int_{{\bf r}_i}^{{\bf r}_j} {\bf A}({\bf r}) \cdot
d{\bf r}$, with the superconducting flux quantum $\Phi_0 = hc/2e$, denotes
the presence of an external magnetic field produced by the vector potential
${\bf A}({\bf r})$.
The interaction term in Eq.~(\ref{eq:ham}),
\begin{equation}
{\cal H}_{\rm int} = J\sum_{<ij>}\left({\bf S}_i \cdot
{\bf S}_j - \frac{1}{4}n_i n_j \right),
\label {eq:intham}
\end{equation}
is chosen to generate both antiferromagnetic and d-wave superconducting orders
at the mean field level.
Here the spin operator is $S^a_i = c_{i\alpha}^{\dag}\sigma^a_{\alpha\beta}
c_{i\beta}$, where the $\sigma^a$ are Pauli matrices, and the density
operator is 
$n_{i\alpha}= c_{i\alpha}^{\dag}c_{i\alpha}$ with $n_i = n_{i \uparrow} +
n_{i \downarrow}$. An onsite repulsion term of the form $U \sum_i n_{i \uparrow}
n_{i \downarrow}$ can be added to ${\cal H}_{\rm int}$ to ensure the absence
of any onsite s-wave pairing, however, we found that it is not necessary because
such induced s-wave pairing is vanishingly small for our parameter regime.
The chemical potential part of the Hamiltonian,
${\cal H}_{\mu} = - \mu \sum_i n_i$, is used to fix the total density of the system to
the desired value. 

It is commonly believed that this simple model, in the
limit of no double occupancy of any site, is adequate 
to describe
the strongly correlated cuprates at low temperatures.  A mean field treatment
of this model, while missing the strong correlation effects, may still usefully
describe the interplay between static antiferromagnetic order and d-wave
superconductivity \cite{tremblay}.

A mean field decomposition of the above model leads to the following
effective Hamiltonian
\begin{eqnarray}
{\cal H}_{{\rm eff}} &=& -\sum_{i,\delta,\alpha} \left(t + \tau_{i\sigma}^{\delta}\right)
e^{i\phi_{ij}}c_{i\alpha}^{\dag} c_{i+\delta\alpha}\nonumber \\
&-&\sum_{i,\alpha} \left( \mu_i-\eta_i+
\alpha e^{i {\bf Q \cdot r}_i} \sum_{\delta} m_{i+\delta} \right) n_{i\alpha}\nonumber \\
&+& \sum_{i,\delta}\left( \Delta_i^{\delta}
[c_{i+\delta\uparrow}^{\dag} c_{i\downarrow}^{\dag} +
c_{i\uparrow}^{\dag} c_{i+\delta\downarrow}^{\dag}] +{\rm h.c.}\right)
\label {eq:Heff}
\end{eqnarray}
where $\tau_{i\sigma}^{\delta}$ is the Fock shift that renormalizes the 
hopping amplitude $t$ and $\eta_i$ is the Hartree shift that renormalizes the
chemical potential $\mu$. 
Here $\delta = \pm \hat x, \pm \hat y$, the four nearest neighbours of any site,
and $\phi_i^{\delta}$ is same as $\phi_{ij}$ with ${\bf j} = {\bf i} +
{\bf \delta}$.
In the above equation the local dSC ($\Delta_i^{\delta}$) and AFM ($m_i$) order
parameters, and the Hartree ($\eta_i$) and Fock ($\tau_{i\sigma}$) shifts
satisfy the following self-consistency conditions:
\begin{eqnarray}
\Delta_i^{\delta} &=& \frac{J}{4} \left( \langle c_{i+\delta\downarrow}
c_{i\uparrow} \rangle + \langle c_{i\downarrow} c_{i+\delta\uparrow}\rangle
\right) \nonumber \\
m_i &=& \frac{J}{2} \left( \langle n_{i\uparrow} \rangle -
\langle n_{i\downarrow} \rangle \right) e^{i {\bf Q \cdot r}_i}\nonumber \\
\tau_{i\sigma}^{\delta} &=& \frac{J}{2} \langle c_{i+\delta,-\sigma}^{\dag}
c_{i,-\sigma} \rangle \nonumber\\
\eta_i &=& - \frac{J}{4}\sum_{\delta}\langle
n_{i+\delta} \rangle
\label {eq:DMdefn}
\end{eqnarray}
where ${\bf Q}=(\pi,\pi)$ is the antiferromagnetic wave vector.
The Hamiltonian in Eq.~(\ref{eq:Heff}) has been studied for a range of model
parameters. In the following we will report results primarily from our
calculations at zero temperature $(T=0)$ with parameters $J/t=1.15$ and
$\langle n \rangle = 0.875$.
In the absence of an external magnetic field, these parameters result in a
uniform d-wave gap, $\Delta_{{\rm max}}\sim 0.36t$ (which is $8~\Delta^{\delta}$
in our notation), with superconducting coherence
length $\xi_0 \approx 4a$ (where $a$ is the lattice spacing), a value consistent
with high $T_c$ cuprates.

We have also studied an extended Hubbard model where the dSC is generated by a
nearest neighbour attractive interaction of the form:
${\cal H}_{\rm int} = -J' \sum_{<ij>,\alpha,\alpha'} n_{i\alpha} n_{j\alpha'}$.
Self consistent calculation on such a model will generate a triplet
superconducting order in addition to a singlet dSC order when decomposed in a
mean field manner. Also, such an extended Hubbard interaction does not by itself
generate AFM order, which has to be incorporated into the model using an
additional interaction parameter \cite{ting1}. By contrast, both dSC and AFM
orders are generated from the single $J$-term in Eq.~(\ref{eq:intham}) in our
calculations, thereby putting both orders on an equal footing. Another
problem of the extended Hubbard model is that, the strength of the onsite
repulsion, $U$, required to generate AFM order has to be fine-tuned to study the
interplay of the two competing orders. Non-zero AFM order develops only
for $U \geq U_c$, a result consistent with earlier findings~\cite{ting1}. On
the other hand a strong $U$ would generate strong antiferromagnetism that would
in turn kill d-wave superconductivity. Except for these limitations, if the triplet
order is {\it ignored}, the mean field decomposition of the extended Hubbard
model generates an effective Hamiltonian similar to that given by
Eq.~(\ref{eq:Heff}), but with different values for the Hartree and Fock shifts.
We have checked that the qualitative physics described by the extended Hubbard
model at the mean field level remains similar to that for Eq.~(\ref{eq:Heff}),
reported in this work.

${\cal H}_{{\rm eff}}$ can be diagonalized using Bogoliubov tranformations,
\begin{equation}
c_{i\alpha} = \sum_n \left\{ u_{n,\alpha}(i) \gamma_{n,\alpha}-\alpha
v_{n,\alpha}^{*}(i) \gamma_{n,-\alpha}^{\dag}\right\}
\label {eq:tranf}
\end{equation}
leading to the usual
BdG equations
\begin{equation}
\left(\matrix{\hat\xi-\alpha\hat M & \hat\Delta \cr \hat\Delta^{*} &
-\hat\xi^{*}-\alpha\hat M} \right)
\left(\matrix{u_{n,\alpha} \cr v_{n,-\alpha}} \right) = E_{n}
\left(\matrix{u_{n,\alpha} \cr v_{n,-\alpha}} \right)
\label {eq:bdg}
\end{equation}
where
$\hat\xi u_{n,\alpha}(i) = -\sum_{\delta}\tilde{t_i^{\delta}} u_{n,\alpha}
(i+\delta)  -\tilde{\mu_i} u_{n,\alpha}(i)$, $\hat M u_{n,\alpha}(i) = \left(
\sum_{\delta} m_{i+\delta} \right)
e^{i{\bf Q \cdot r_i}} u_{n,\alpha}(i)$ and
$\hat\Delta u_{n,\alpha}(i) = \sum_{\delta}(\Delta_i^{\delta}+
\Delta_{i+\delta}^{-\delta}) u_{n,\alpha}(i+\delta)$.
One can numerically diagonalize Eq.~(\ref{eq:bdg}) directly.  However,
exploiting the magnetic translation symmetry of the effective
Hamiltonian allows one to treat much larger system sizes.  This
procedure was discussed by Wang and MacDonald\cite{wang}.  However,
since our model and resulting BdG equations differ somewhat, we describe
below the block diagonalization procedure we use.

We assume that our system is made up of a periodic array of identical
magnetic unit cells of size $N_x \times N_y$. For our case we take
$N_x=N_y/2$, a rectangular magnetic unit cell enclosing two superconducting
flux quanta, and  work in the Landau gauge,
${\bf A}({\bf r}) = (0,Hx)$. 
In this gauge the magnetic translation 
operators are $\langle {\bf r}|{\cal T}_R|{\bf r}' \rangle = \delta_{{\bf r},{\bf
r+R}} e^{-ibR_x(i_y+R_y)}$, where the parameter $b$ is defined as
$b \equiv H/\Phi_0 = 2\pi/N_xN_y$.
Since $[{\cal H}_{{\rm eff}}, {\cal T}_R]=0$, 
the eigenstates of ${\cal T}_R$ can be used to block diagonalize
${\cal H}_{{\rm eff}}$.  The
transformations that block diagonalize the BdG equations in the magnetic
wave vector, ${\bf k}$, are
\begin{eqnarray}
u_{i,\alpha}^n({\bf R}) &=& e^{i {\bf k \cdot R}} e^{-ib(i_y+R_y)R_x}
u_{i,\alpha}^n({\bf k})\nonumber \\
v_{i,\alpha}^n({\bf R}) &=& e^{i {\bf k \cdot R}} e^{ib(i_y+R_y)R_x}
v_{i,\alpha}^n({\bf k})
\label{eq:block}
\end{eqnarray}
where $k_x=\frac{2\pi m_x}{p N_x}$, with $m_x=0,1,2,...,(p-1)$ and
$k_y=\frac{2\pi m_y}{q N_y}$, with $m_y=0,1,2,...,(q-1)$. 
The eigenvalue problem for a system of size
$pN_x \times qN_y$ reduces to $p \times q$ eigenvalue matrix equations,
each for a system of size $N_x \times N_y$.
The block diagonal BdG matrix of Eq.~(\ref{eq:bdg}) becomes, for each ${\bf k}$,
\begin{eqnarray}
\left(\matrix{\hat\xi_{\alpha}({\bf k})-\alpha\hat M({\bf k}) & \hat\Delta
({\bf k}) \cr \hat\Delta^{*}({\bf k}) &
-\hat\xi^{*}_{\alpha}({\bf k})-\alpha\hat M({\bf k})} \right)
\left(\matrix{u_{i,\alpha}^n({\bf k}) \cr v_{i,-\alpha}^n({\bf k})} \right)
\nonumber \\
=
E_{\alpha}^n({\bf k})
\left(\matrix{u_{i,\alpha}^n({\bf k}) \cr v_{i,-\alpha}^n({\bf k})} \right)\ .
\label {eq:bdgK}
\end{eqnarray}
Here the operators (in the bulk) are defined as 
\begin{eqnarray}
\hat\xi_{\alpha}({\bf k})u_{i,\alpha}^n({\bf k}) &=& - \sum_{\delta}
\left\{t+\frac{J}{2} \langle c_{i+\delta,-\alpha}^{\dag}
c_{i,-\alpha} \rangle \right\}e^{i\phi_i^{\delta}}
u_{i+\delta,\alpha}^n({\bf k})\nonumber \\
\hat M({\bf k}) u_{i,\alpha}^n({\bf k}) &=& \left\{\sum_{{\bf q},n'}
m_{{\bf q}n'}(i)\right\} u_{i,\alpha}^n({\bf k})\nonumber \\
\hat\Delta({\bf k}) v_{i,-\alpha}^n({\bf k}) &=& 2\sum_{\delta} \left\{
\sum_{{\bf q},n'} \Delta_{{\bf q}n'}^{\delta}(i) \right\} v_{i+\delta,-\alpha}
^n({\bf k})\ .
\label{eq:Kdefn}
\end{eqnarray}
In this `repeated zone scheme' the self-consistency conditions,
Eq.~(\ref{eq:DMdefn}), can be rewritten as
\begin{eqnarray}
\Delta_{{\bf q}n}^{\delta}(i) &=& \frac{J}{4{\cal N}} \left[\left\{
u_{j,\downarrow}^n({\bf q})v_{i,\uparrow}^{n*}({\bf q})+
u_{i,\downarrow}^n({\bf q})v_{j,\uparrow}^{n*}({\bf q}) \right\}
(1-f_{{\bf q} n\downarrow}) \right. \nonumber \\
&+&\left. \left\{u_{i,\uparrow}^n({\bf q})v_{j,\downarrow}^{n*}({\bf q})+
u_{j,\uparrow}^n({\bf q})v_{i,\downarrow}^{n*}({\bf q}) \right\}
f_{{\bf q} n\uparrow} \right] \\
m_{{\bf q}n}(i) &=& \frac{U}{2{\cal N}}e^{i {\bf Q \cdot r}_i} \left[
|u_{i,\uparrow}^n({\bf q})|^2 f_{{\bf q} n\uparrow} -
|u_{i,\downarrow}^n({\bf q})|^2 f_{{\bf q} n\downarrow} \right. \nonumber \\
&+&\left. |v_{i,\uparrow}^n({\bf q})|^2 (1-f_{{\bf q} n\downarrow}) -
|v_{i,\downarrow}^n({\bf q})|^2 (1-f_{{\bf q} n\uparrow}) \right]
\label {eq:SCdm}
\end{eqnarray}
where $j=i+\delta$ and ${\cal N}=p\times q$, the total number of magnetic unit
cells.  Note that, for our choice of gauge, the matrices
$\hat\xi_{\alpha}({\bf k})$,$\hat M({\bf k})$ and
$\hat\Delta({\bf k})$ are ${\bf k}$-independent for the bulk of the
system, as can be seen from
 Eq.~(\ref{eq:Kdefn}). However 
${\bf k}$-dependence enters through the boundary terms (See Appendix). 

Starting with an initial guess for all {\it local} variables (dSC order, AFM order
and Hartree and Fock shifts) and for the value of $\mu$, 
we solve for all eigenvalues and eigenvectors of the BdG
matrix for each ${\bf k}$, in terms of which all local orders are
recalculated using self-consistency conditions Eq.~(\ref{eq:SCdm}).
This process is repeated until a convergence is obtained in these variables and
also in $\mu$ which fixes the total density to the desired value of $0.875$.

\section{Results}\label{sec:results}

\subsection{Spatial structure of local order parameters}
\label{subsec:op}

In this section we discuss the structure of the dSC, AFM and charge density 
orders, as well as of the $\pi$-triplet SC order which is induced in the
presence of coexisting AFM and dSC orders.

\vskip 0.3cm

\begin{figure}
\vskip0mm
\hspace*{0mm}
\psfig{file=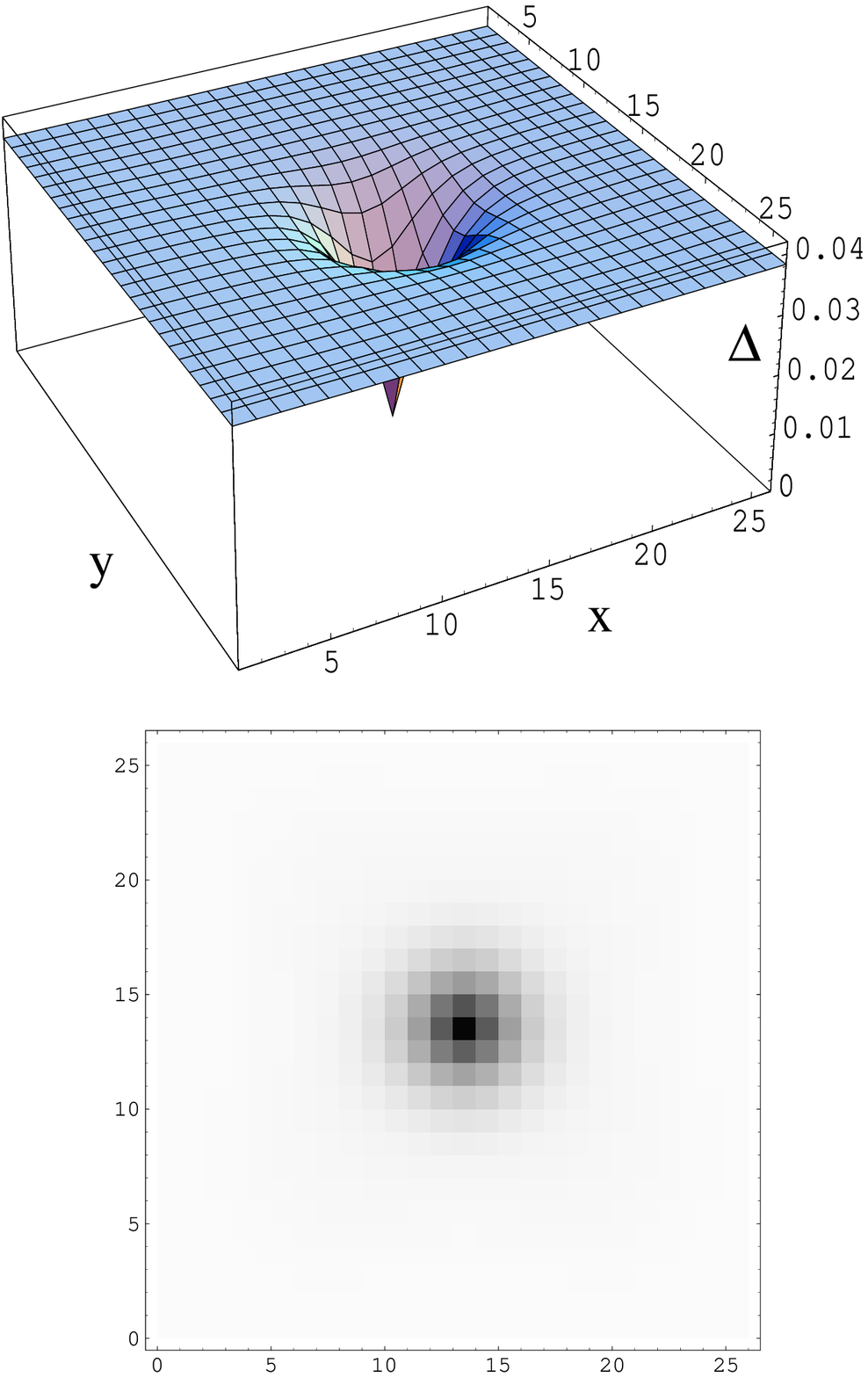,width=3.3in,angle=0}
\vskip3mm
\caption{
Upper panel: Singlet d-wave pairing amplitude $\Delta_i$ on a magnetic unit cell of
size $N=26\times26$, with $J=1.15t$ and $\langle n\rangle=0.875$.
$\Delta_i$ is uniform (and equal to the BCS value)  away from the vortex core,
and falls to zero at the vortex center with the length scale $\xi$.\\
Lower panel: The grey-scale plot shows the length scale $\xi$. The dark (light)
regions indicate smaller (larger) values for $\Delta_i$.
Note that only {\it half} of the unit cell, containing one flux quantum, is
shown here (and in all other figures). The full unit cell
configuration can be obtained by periodically repeating the results along
$\hat{y}$-direction.
}
\label{fig:del}
\end{figure}

\noindent
{\bf d-wave Pairing Amplitude:}
For our lattice BdG calculation we defined a d-wave pairing amplitude
$\Delta^{\delta} ({\bf i},{\bf R})$ following Eq.~(\ref{eq:DMdefn}) that lives
on bonds. We can
also define a gauge invariant site dependent singlet d-wave order parameter in
our model as,
\begin{equation}
\Delta_i = \frac{1}{4}\left[\Delta^{\hat x}({\bf r})+
\Delta^{-\hat x} ({\bf r}) -e^{-i b x} \Delta^{\hat y}({\bf r})-e^{i b x}
\Delta^{-\hat y}({\bf r}) \right]
\label{eq:delr}
\end{equation}
where $r\equiv ({\bf i},{\bf R})$. In the left panel of
Fig.~(\ref{fig:del}) we plot the magnitude of $\Delta_i$ on half of a magnetic
unit cell.  The spatial dependence of $\Delta_i$ is very similar to its behavior
in the absence of AFM order (not shown here).  The size of the vortex core, $\xi$,
is similar for the two cases, although the initial slope inside the core is
somewhat smaller in the case with AFM order.  This is what one would expect within
Ginzburg-Landau theory~\cite{john,huzhang}.

The presence of the vortex lattice induces an {\it extended} s-wave order
($\Delta_i^s$) around vortex in a self-consistent manner; defined as:
\[ \Delta_i^s = \left[\Delta^{\hat x}({\bf r})+\Delta^{-\hat x} ({\bf r})+e^{-i b x}
\Delta^{\hat y}({\bf r})+e^{i b x}\Delta^{-\hat y}({\bf r}) \right]/4. \]
However, the strength of the induced $\Delta^s$ order is vanishingly small
compared to the primary d-wave order, and hence it will be ignored in further
discussions.

\vskip 0.3cm

\noindent
{\bf Antiferromagnetic Order:}
In the absence of an external magnetic field, the antiferromagnetic order is
stabilized at half filling ($\langle n \rangle=1$) and away from half filling,
AFM order is quickly suppressed.

\noindent
For our choice of parameters, the minimum
energy configuration is one in which  AFM order is identically zero for the uniform
system at zero external magnetic field ($H=0$). However, at low temperatures, 
in the presence of a magnetic field, AFM order develops self-consistently
in the vicinity of the vortex cores where superconductivity is destroyed.
The structure of the AFM order is very
different from that predicted earlier \cite{ting1}. We have confirmed that the
structure of our order parameter, which changes sign within the unit cell as is
discussed below, minimizes the ground state energy, $E_g$, the expectation value
of the Hamiltonian of Eq.~(\ref{eq:ham}), evaluated in the mean field ground state.

\begin{figure}
\vskip-2mm
\hspace*{0mm}
\psfig{file=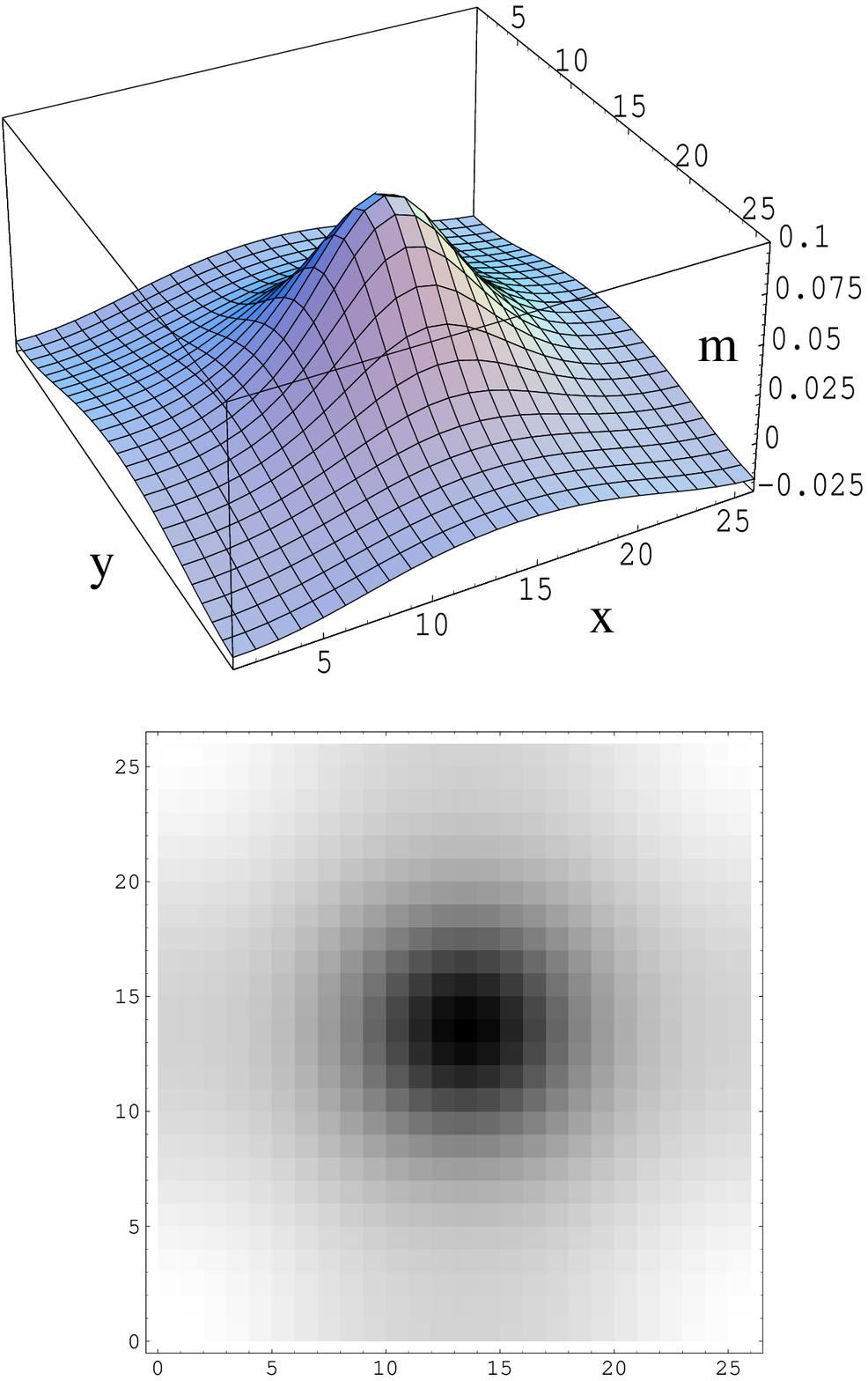,width=3.2in,angle=0}
\vskip3mm
\caption{
Upper panel: Staggered magnetization $m_i$ on a magnetic unit cell of size
$N=26\times26$ system, with $J=1.15t$ and $\langle n\rangle=0.875$.
$m_i$ is strongest in the vortex cores, and decays very slowly away from it.\\
Lower panel: The grey-scale plot shows the much larger length
scale for the decay of $m_i$ compared to $\xi$ (See also Fig.~\ref{fig:del}).
The dark (light) regions indicate larger (smaller) values for $m_i$.
}
\label{fig:mag}
\end{figure}
In this regard, we
emphasize the role of including the Hartree and Fock shift terms. In a
variational calculation, such as the BdG method, it is possible to achieve
self-consistency for various different order parameter configurations
corresponding to different $E_g$'s. The correct configuration is the one that
minimizes $E_g$. Ignoring the Hartree and Fock shifts will give a higher value
for  $E_g$.

The order parameter configuration in Ref.~\cite{ting1}, when
periodically repeated for an array of unit cells along $\hat x$ and $\hat y$
directions, produces an AFM order that changes sign on alternate unit cells
along the $\hat y$-direction, and has the same sign along $\hat x$-direction.
Such a ``stripe-like" AFM order spontaneously breaks the symmetry of the $\hat x$ 
and $\hat y$ directions.
Working with a unit cell containing 4 vortices, we have
verified that such a ``stripe-like" AFM order does not converge to a
self-consistent solution within our Hartree-Fock, BdG formalism. On the other hand,
a ``checkerboard" pattern of AFM order, in which the sign of AFM order 
alternates in adjacent unit cells along {\it both} the $\hat x$ and $\hat y$
directions, leads to a self-consistent solution that has slightly higher $E_g$
than for the periodic configuration reported here.

Comparison of Fig.~(\ref{fig:mag}) to the result for the d-wave order parameter shown
in Fig.~(\ref{fig:del}) above, clearly shows that the length scale of decay 
of the AFM order away from vortex core is much larger than the vortex core size $\xi$,
as one would expect in the regime where uniform dSC is stable in the absence of an
applied magnetic field. 
\begin{figure}
\vskip0mm
\hspace*{0mm}
\psfig{file=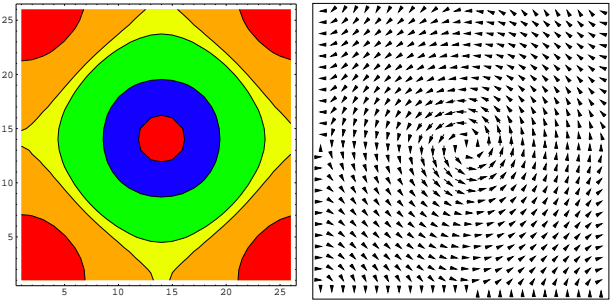,width=3.2in,angle=0}
\vskip3mm
\caption{
Contour plot of staggered magnetization $m_i$ (left panel) on a magnetic unit
cell of size $N=26\times26$ system, which shows that $m_i$ changes sign
creating a domain wall along the line on which $\nabla \times j_s \approx 0$.
The contours correspond to $m=0.075, 0.03, 0.004, 0, -0.01$ from the
center to radially outwards.
Here, $j_s$ is the supercurrent associated with dSC order, which is plotted in
the right panel. Notice that along the line where $m_i \approx 0$, $j_s$ arrows
do not bend, showing that $\nabla \times j_s \approx 0$ along those lines.
}
\label{fig:mag2}
\end{figure}
We also notice in Fig.~(\ref{fig:mag2}) the surprising result that the
staggered magnetization changes sign within the vortex unit cell. Specifically 
it changes sign along the line where the supercurrent $j_s$ associated with the
dSC order parameter has a zero curl, i.e. the line along which $j_s$ changes
the direction of
its winding. This result is true for a wide range of parameters studied (i.e.
$J$, $\langle n \rangle$ and system size or external field strength); as well
as for the study with the extended Hubbard model described in
Sec.~{\ref{sec:model}). This puzzling effect can be thought of as the result of
an effective interaction of the following form:\cite{zhangcomm}
$({\bf \nabla} \times {\bf j})_I({\bf \nabla} \times {\bf j})_J m_I m_J$, where
$I$ and $J$ label adjacent plaquettes and $m_I$ is a course-grained AFM order
associated with the plaquette. Alternatively, this implies that the gradient term 
for the AFM order
in the G-L free energy takes the form $|(\nabla - a)m|^2$, where
$a$ is proportional to the supercurrent.  Note, that this 
interaction implies  no preferred
sign for the AFM order, but only that it changes sign along the line where the
supercurrent vorticity changes sign.  We have confirmed that this is also
the case for our numerical BdG solution, i.e. there is no preferred direction for 
the AFM order. 

\vskip 0.3cm

\noindent
{\bf Charge Density Order:}
In the absence of AFM order the charge density is relatively structureless; 
$\langle n_i \rangle$
is uniform everywhere except very near a vortex. Around the vortex it has a
very weak dip which changes to a weak peak at the vortex center.
We believe that such weak
density fluctuations would not survive long range Coulomb screening effects and
hence will not have any important effects in a more realistic model. However,
when the AFM order is allowed to develop self consistently, the charge
density reorganizes itself significantly to accommodate the AFM order, as shown
in Fig.~(\ref{fig:den}). 
\begin{figure}
\vskip-1mm
\hspace*{0mm}
\psfig{file=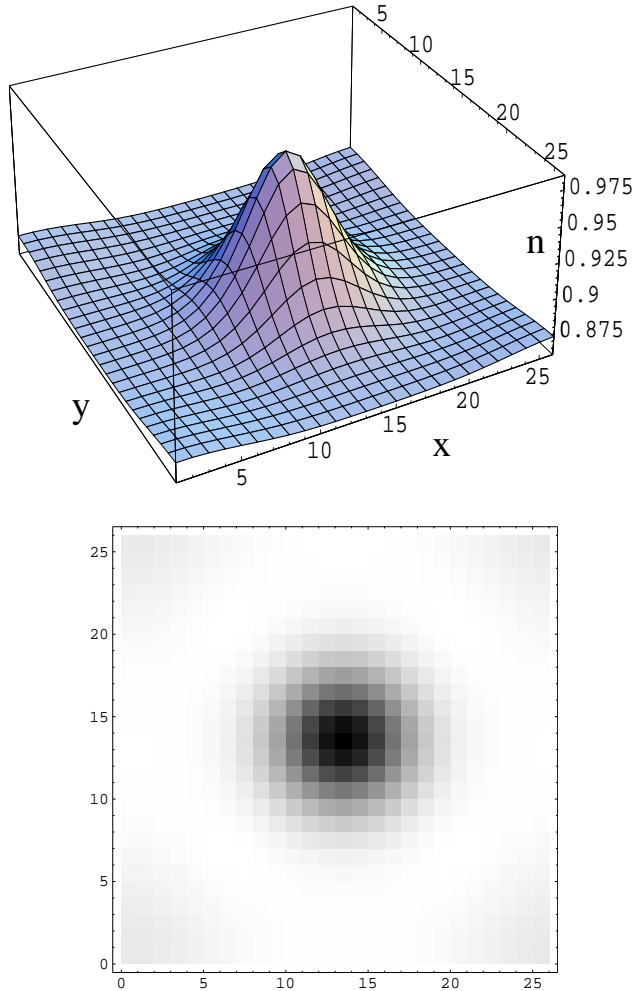,width=3.3in,angle=0}
\vskip3mm
\caption{
Upper panel: Charge density $\langle n_i \rangle$ on a magnetic unit cell of size
$N=26\times26$ system, with $J=1.15t$ and $\langle n\rangle=0.875$.
$\langle n_i \rangle$ is more or less uniform away from the vortex core,
but achieves a value very close to half-filling ($\langle n_i \rangle=1$) at
the vortex
center. This local half-filling at the vortex core favors the AFM
order in the core region where dSC order is depleted due to external magnetic
field.\\
Lower panel: The grey-scale plot shows that the distance from the
vortex center where $\langle n_i \rangle$ differs significantly from its
uniform value is slightly larger than $\xi$. (See also Fig.~\ref{fig:del}.)
The dark (light) regions indicate larger (smaller) values for
$\langle n_i \rangle$.
}
\label{fig:den}
\end{figure}
The local electron density remains more or less uniform away from the vortex
center, but near
the vortex core the magnitude of $\langle n_i \rangle$ increases continuously
resulting in a density close to half filling at the vortex center.
For the $t-J$ model at $T=0$, AFM order becomes the dominant order at half filling
in the absence of the magnetic field. The charge density thus organizes itself
to favor the AFM order strongly at the vortex core by approaching
unity. Again, our result is different from
that in Ref.~\cite{ting1}, where $\langle n_i \rangle$ differs from the
uniform bulk value only within one lattice spacing of the vortex center. 
Full self-consistency, including the Hartree and Fock terms, is required to
reliably study charge inhomogeneities.  However, one does need to keep in mind
that the long-range Coloumb interaction in a three-dimensional material 
can further alter the charge
distribution in ways which are beyond the scope of the present study.

Other self-consistent local variables such as the Hartree and Fock shift
terms (not shown here) also become weakly inhomogeneous  around the vortices.

\vskip 0.3cm

\noindent
{\bf $\pi$-triplet Order Parameter:}
In the presence of coexisting dSC and AFM orders, it had been argued
\cite{bamsoo,murakami} that an additional order, called $\pi$-triplet order,
will develop self-consistently in the absence of any external magnetic
field. It was shown using mean-field gap-equations that even in the absence 
of any interaction that generates $\pi$-triplet pairing directly, this order will
be induced by coexisting DSC and AFM order. We define local $\pi$-triplet order
as:
\begin{equation}
\Pi_i^{\delta}=\frac{J}{4} \left\{\langle c_{i+\delta\downarrow}
c_{i\uparrow} \rangle - \langle c_{i\downarrow} c_{i+\delta\uparrow}\rangle
\right\}e^{i {\bf Q \cdot r}_i}.
\label{eq:trip}
\end{equation}
This defines a triplet pairing amplitude
in the $S_z=0$ channel.  In the uniform
system, Fourier transformation of $\Pi_i^{\delta}$ results in $\langle
c_{{\bf -k}\downarrow} c_{{\bf k+Q}\uparrow} \rangle$, so that the center of
mass momentum for the pair is ${\bf Q}=(\pi,\pi)$, and hence the name. (The
prefactor $J/4$ is arbitrary in the absence of any microscopic interaction
that generates this order, we use this prefactor to compare the strength
of this order relative to $\Delta_i^{\delta}$.) One physical way of understanding
the induction of the $\pi$-triplet order is the folowing. The dSC quasiparticles
carry momenta (${\bf k}\uparrow$, -${\bf k}\downarrow$), whereas the AFM
quasiparticles carry momenta (${\bf k+Q}\uparrow$, ${\bf k}\uparrow$). As a result,
in the presence of both dSC and AFM order, there is a possibility of pairing
between quasiparticles with momenta (${\bf k+Q}\uparrow$, -${\bf k}\downarrow$)
which leads to the $\pi$-triplet pairing.

Our $t-J$ Hamiltonian does not contain any interaction involving $\Pi_i^{\delta}$
at the mean field level (Eq.~\ref{eq:Heff}), but the BdG self-consistency
induces $\Pi_i^{\delta}$ around the vortex, as shown in Fig.~(\ref{fig:trip}).
On the other hand, the extended
Hubbard model supports $\pi$-triplet pairing, and such an interaction should be
present in the microscopic mean-field Hamiltonian equivalent to
Eq.~(\ref{eq:Heff}).
We have verified that the qualitative results from the extended Hubbard model
are similar to the results presented here.
\begin{figure}
\vskip0mm
\hspace*{0mm}
\psfig{file=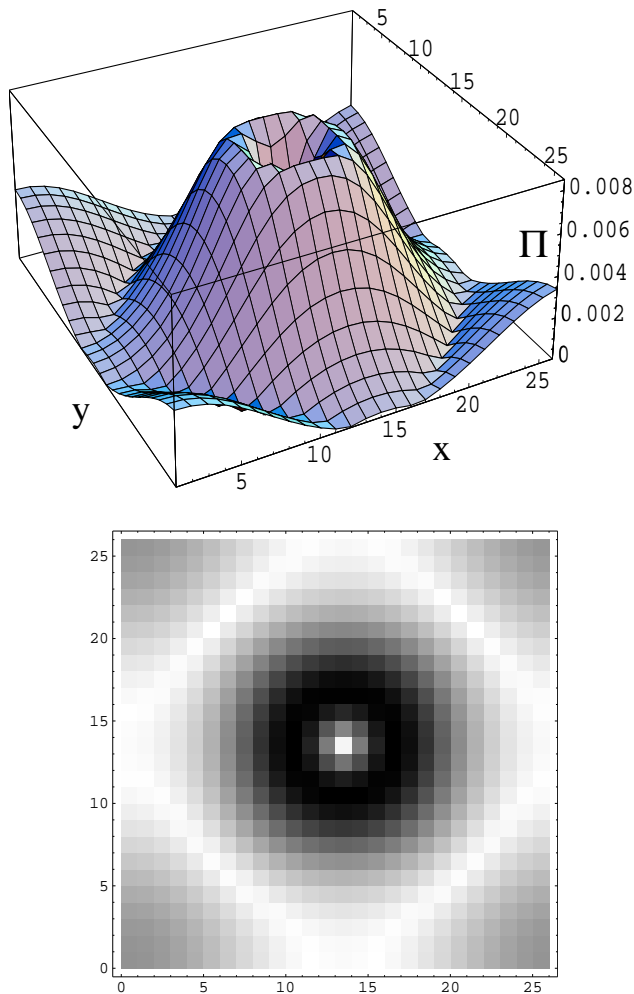,width=3.3in,angle=0}
\vskip3mm
\caption{
Upper panel: $\pi$-triplet pairing amplitude $\Pi_i$ induced on a magnetic unit cell of
size $N=26\times26$ system, with $J=1.15t$ and $\langle n\rangle=0.875$.
$\Pi_i$ is zero at the vortex core, rises around the vortex,
and falls gradually at larger distances.\\
Lower panel: The grey-scale plot for $\pi$-triplet showing the length scale of
decay which is similar to the AFM order (in Fig.~\ref{fig:mag}).
The dark (light) regions indicate larger (smaller) values for $\Pi_i$.
$\Pi_i$ also goes to zero on the lines of zero
supercurrent.
}
\label{fig:trip}
\end{figure}

This $\Pi_i^{\delta}$ order parameter was first introduced by Zhang \cite{zhang} 
in a consistent picture for AFM and dSC order for cuprates using SO(5) theory.
It was shown using general symmetry arguments that a theory which allows
$\Delta_i^{\delta}$, $\Pi_i^{\delta *}$ and $m_i$ orders will satisfy the
relation:
\begin{equation}
\tilde{\Delta_i^{\delta}} \tilde{\Pi_i^{\delta *}} = - \tilde{m_i}
(1-\langle n_i \rangle),
\label{eq:so5reln}
\end{equation}
where the variables with `tilde' are the same as in Eq.~(\ref{eq:DMdefn}) and
in Eq.~(\ref{eq:trip}), but {\it without} the prefactors involving the coupling
strength.
This relation can be understood using a simpler picture of a so-called
``spin-flop'' model of an easy-axis AFM in a parallel magnetic field.
Zhang~\cite{zhang} discussed the analogy between his SO(5) model and the
spin-flop model, where for small fields the equilibrium state is one with
an AFM order parameter, $N_z$, along the easy axis.  As the field is
increased above the spin-flop transition, the Neel vector flops into
the plane with a value $N_\perp$ and there is also a uniform magnetization,
$M_z$, along the field.  The case in which dSC and AFM order coexist
corresponds to one in which the Neel vector $\vec N = (N_\perp,N_z)$ is
tipped out of the plane and the uniform magnetization $\vec M = (M_\perp,M_z)$ 
is tipped away from the easy axis.  $M_\perp$ is analogous to the $\Pi$
triplet order parameter.  Since $\vec N$ is the difference of the two 
sublattice magnetizations, while $\vec M$ is their sum, $\vec N \cdot
\vec M = 0$.  Using the correspondence between the two models, 
$m \leftrightarrow N_z,\ \Delta \leftrightarrow N_\perp ,
(1-\langle n_i \rangle)\ \equiv\ n_H
\leftrightarrow M_z,\ \Pi \leftrightarrow M_\perp$, this implies the
relation of Eq.~(\ref{eq:so5reln}).

As seen from Fig.~(\ref{fig:den}), $\langle n_{{\rm vortex}} \rangle \approx
0.975$ at the
vortex center and hence does not achieve a value of true half-filling. However,
calculations with increasing system sizes indicate that with larger system
sizes the value approaches closer to $\langle n_i \rangle = 1$. Defining the
local hole density as ($\langle n_{{\rm vortex}} \rangle - \langle n_i \rangle$),
instead of ($1- \langle n_i \rangle$), we found that Eq.~(\ref{eq:so5reln}) is
satisfied within $5\%$ accuracy. Although it is not obvious that our model has
the same symmetry as the spin-flop model, we nevertheless find that this
non-trivial relationship (Eq.\ref{eq:so5reln}) holds rather well
{\it site by site}!

\subsection{Local Density of States (LDOS) at the vortex core}
\label{subsec:ldos}

Having understood the self-consistent spatial structures of the relevant order
parameters in an antiferromagnetic d-wave vortex, we turn our attention to the
local density of states around the core. Study of the LDOS provides
information about the low energy quasiparticle structure which is responsible for
most of the novel phenomena associated with the cuprates. Also the LDOS is
proportional to the tunneling conductance that is measured directly in STM
experiments \cite{pan,hoffman} and hence can provide a direct test for the validity
of theoretical predictions. As discussed above, a pure d-wave BCS mean
field calculation \cite{wang} produces a large,unobserved ZBCP in the LDOS at the core
of a d-wave vortex lattice. The ZBCP can be understood as the effect of
Andreev reflection of d-wave quasiparticles within the vortex core. 
Andreev reflection is caused by the rotation of the internal state of an
excitation from particle-type to hole-type (or vice versa) \cite{andreev}. For a d-wave
order parameter, such reflections scramble the phase information of the order parameter,
and, as a result, the coherence peaks of the dSC collapse onto a ZBCP in the LDOS at
the vortex core. Though the destruction of coherence peaks will occur for a wide
range of angle for a reflecting surface, the effect is most drastic for the
angle of $\pi/4$. The obvious question is: Does the additional AFM order
suppress the unphysical ZBCP at the vortex core? To address this
issue, we present the LDOS at the vortex core as obtained from our calculations.

The single particle local density of states at $T=0$ is calculated from the expression:
\begin{eqnarray}
N(i,\omega)&=&\frac{1}{{\cal N}} \sum_{n,{\bf k},\alpha} \left\{
|u_{i,\alpha}^n({\bf k})|^2 \delta \left(\omega-E_{\alpha}^n({\bf k})\right)
\right. \nonumber \\
&+&\left.|v_{i,\alpha}^n({\bf k})|^2 \delta \left(\omega+E_{\alpha}^n({\bf k})\right)
\right\}
\label {eq:lDOS}
\end{eqnarray}
where the individual delta functions are broadened with a width comparable to
the average energy level spacing. The results are shown in Fig.(\ref{fig:ldos}). 
\begin{figure}
\vskip-2mm
\hspace*{0mm}
\psfig{file=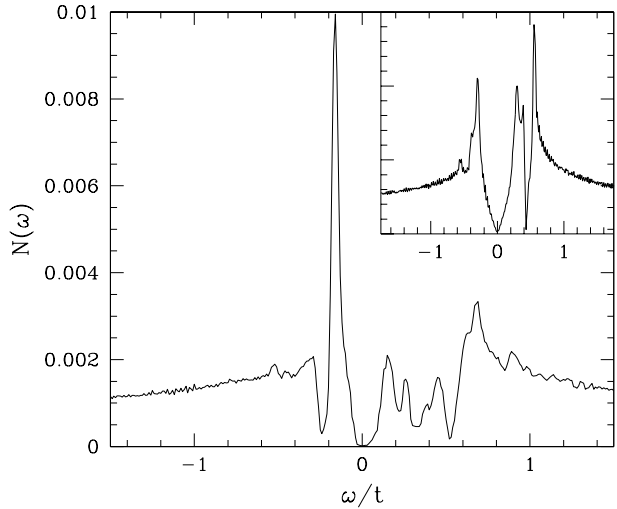,width=3.2in,angle=0}
\vskip3mm
\caption{Normalized density of states $N(\omega)$ at the vortex core for
$J=1.15t$ and $\langle n\rangle=0.875$.
The competition of dSC and AFM order splits the zero bias
peak which would be present in the absence of AFM order. The signature of 
a spin gap appears at large $\omega \sim 0.5$.
The inset shows the DOS for a uniform system in the absence of vortices
($\hat{y}$-axis levels are not given for clarity). The parameters for the inset
are $\Delta_{{\rm max}}=0.36$, $\mu=0.35$ and $m=0.1$, consistent with BdG
results.
}
\label{fig:ldos}
\end{figure}
There are two important features in Fig.(\ref{fig:ldos}) that require
further explanations.
(i) The zero bias peak splits in two, and the coherence peaks are suppressed.
(ii) An extra feature appears at large positive $\omega$ which looks like a
second gap. Also, $N(\omega)$ appears to be quite asymmetric.

The asymmetry and the second gap scale (at $\omega \sim 0.5t$) in the LDOS
can be understood by analyzing the problem of a
uniform d-wave SC coexisting with AFM order in the absence of vortex. This
is described by the Hamiltonian $H=\sum_{{\bf k}}^{\prime} \Psi_{\bf k}^{*}
M_{\bf k} \Psi_{\bf k}$ where,
\begin{equation}
M_{\bf k}=\left(\matrix{\xi_{\bf k}-\mu & \Delta_{\bf k} & -m & \Pi_{\bf k} \cr
\Delta_{\bf k} & -\xi_{\bf k}+\mu & -\Pi_{\bf k} & -m \cr
-m & -\Pi_{\bf k} & -\xi_{\bf k}-\mu & -\Delta_{\bf k} \cr
\Pi_{\bf k} &-m & -\Delta_{\bf k} & \xi_{\bf k}+\mu} \right)
\label {eq:uniform}
\end{equation}
$\Psi_{\bf k} = \left(c_{{\bf k}\uparrow},~~c_{{\bf -k}\downarrow}^{\dag},
~~c_{{\bf k+Q}\uparrow},~~c_{{\bf -k-Q}\uparrow}^{\dag} \right)^{T}$ and the
prime on the ${\bf k}$-sum indicates a sum over only half of the Brillouin zone,
which has been reduced by AFM order. Here
$\xi_{\bf k} = -2t ({\rm cos}k_x+{\rm cos}k_y)$, $\Delta_{\bf k} = \Delta_0
({\rm cos}k_x-{\rm cos}k_y)$ and $\Pi_{\bf k} = \Pi_0 ({\rm cos}k_x-{\rm cos}k_y)$.
The result of the spin-density wave (SDW) problem with $\Delta_0=0$,
$\Pi_0=0$ and $\mu = 0$ is well known; under these conditions diagonalization
of Eq.~(\ref{eq:uniform}) results in a $N(\omega)$ that has a spin gap of size
$m$ 
at the Fermi energy. Inclusion of particle hole asymmetry $(\mu \neq 0)$ shifts
the overall density of states by an energy, $\mu$. Superconductivity can 
now be introduced with a non-zero $\Delta_0$, which will open a superconducting 
gap at $\omega=0$ with d-wave symmetry. The resulting DOS for this uniform system,
which is plotted as an inset of Fig.~(\ref{fig:ldos}),
shows two gap scales, one at $\omega=0$ and the other at
$\omega=|\mu|+m$. With $m=0$, a non-zero $\Pi_0$ in Eq.~(\ref{eq:uniform})
splits the dSC coherence peaks
at $\Delta_{{\rm max}} \pm \Pi_0$; the splitting becomes asymmetric when
$\mu \neq 0$. However, the $\pi$-triplet order does not change $N(\omega)$
qualitatively from that in the inset of Fig.~(\ref{fig:ldos}) (at least for
the choice of our parameters). (See also Fig.~(10) of Ref.~\cite{murakami}.)
The presence of the vortex modifies the LDOS structure near $\omega=0$
(described in the next paragraph), but the SDW gap  at larger $\omega$ survives
the vortex excitations and shows up in Fig.~(\ref{fig:ldos}). The above argumnet
for the origin of the `spin-gap scale' was verified by tuning model parameters
that change $\mu$ and $m$. It was found that the `spin-gap' changes accordingly
in our BdG solutions.

Next we turn our attention to the splitting of the zero bias peak in $N(\omega)$.
We have already argued (See Sec. II) that the ZBCP in the LDOS in Ref.\cite{wang}
is due to the Andreev reflection of d-wave quasiparticles from the vortex core
which acts like a boundary.
The effects of an additional subdominant ordering
on the ZBCP at the surface have been investigated recently
\cite{matsumoto2,sig,rainer,buchh}. The conclusion is that, if the subdominant order breaks 
time-reversal symmetry, then the ZBCP will split in two. For our case, the
subdominant order is AFM order, which breaks time-reversal symmetry, so
that the presence of the AFM order within vortex core should split
the ZBCP, as found in Fig.~(\ref{fig:ldos}). Within this formalism, the energy
difference between the split peaks would be proportional to the value of $m_i$
at the vortex core, which is consistent with our findings.

Having understood the features of the LDOS at the vortex core as found
in our calculations, we must emphasize that, a mean-field calculation on a
model with static antiferromagnetism like
ours, does not reproduce all the experimental findings. Although the competition
between dSC and AFM order produces a gap near the Fermi energy in the DOS -- a
result consistent with experimental findings -- the spin-gap at larger
$\omega$ has not been observed in experiments. Also, the strong split-peak around
$\omega = 0$, which is due to static AFM does not resemble the weak
subgap hump found in tunneling conductance data. The strong split peaks, however,
die off rapidly away from the vortex center, as seen from the top panel of
Fig.~(\ref{fig:ldosav}), which shows the evolution of the LDOS along the
diagonal direction from the vortex. We also find that the coherence peaks
build up near $\omega \sim \Delta_{{\rm max}}$ away from the core, as expected.
Similar behaviour is found along the $\hat{x}$ and $\hat{y}$ directions as well.
\begin{figure}
\vskip-2mm
\hspace*{0mm}
\psfig{file=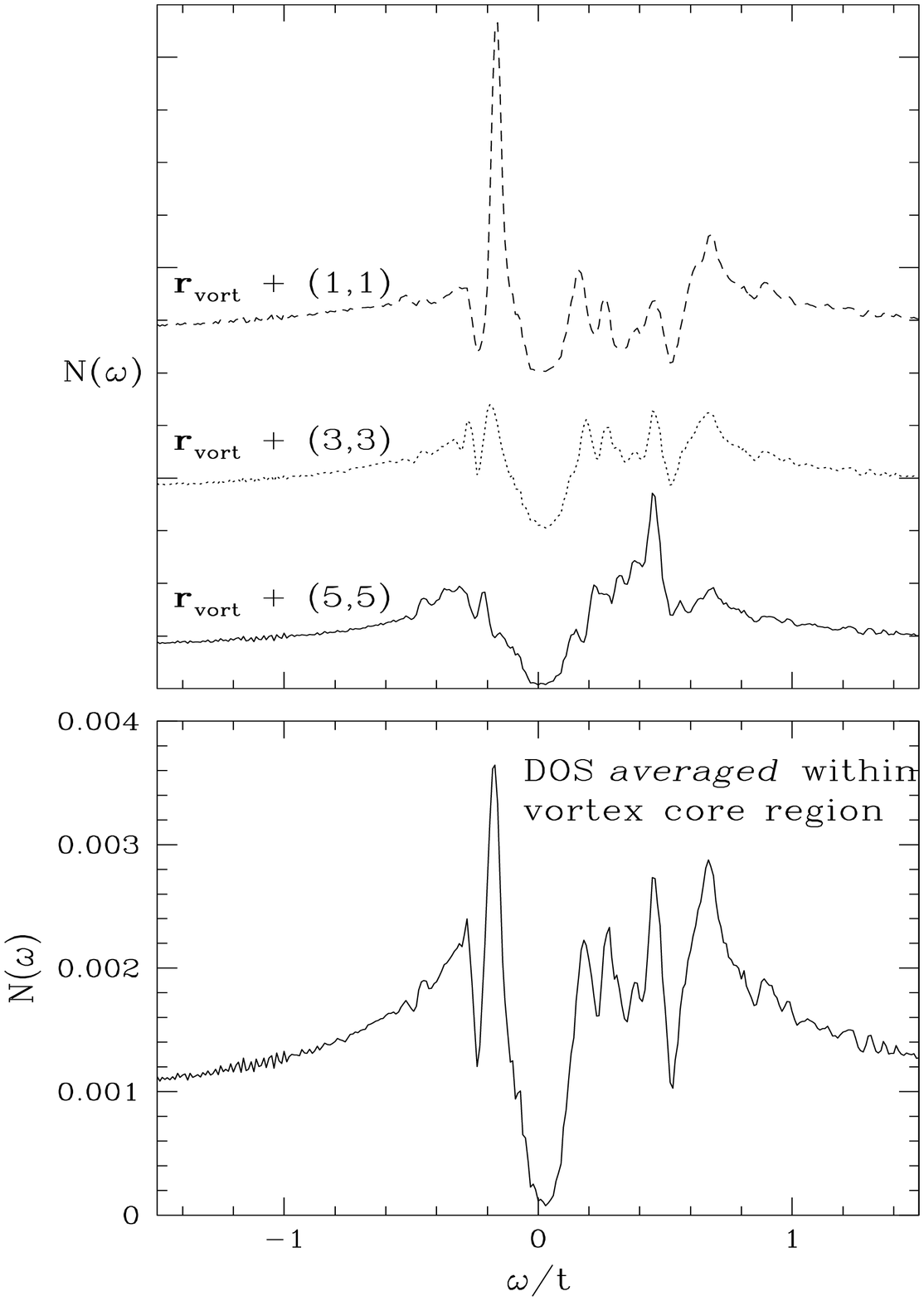,width=3.3in,angle=0}
\vskip0mm
\caption{Upper panel: Local Density of states away from the vortex core along
the diagonal direction. The strong split-peaks fall rapidly in magnitude away
from the vortex center.\\
Lower panel: Normalized local density of states averaged over all sites within
a radius $\xi$ from the vortex center.
}
\label{fig:ldosav}
\end{figure}
Surprisingly the low energy peaks persist to larger distances, {\it even}
outside the vortex core. This is interesting from the experimental point of
view, because such low energy quasiparticle humps have been observed beyond
the vortex core \cite{hoffman}. This led us to calculate the local density of
states {\it averaged} over all sites within a radius $\xi$ from the vortex
core, which is presented in the lower panel of Fig.~(\ref{fig:ldosav}).
Allowing for some uncertainty in locating the vortex center in STM measurements,
we believe the qualitative agreement of our results in
Fig.~(\ref{fig:ldosav}) with experimental findings is significant.

\section{Conclusions}
In conclusion, we have studied the spatial structures of different order parameters
and their effect on the local density of states at the vortex core within a
model that allows for the competition between d-wave superconductivity and
static antiferromagnetism. Our main results are: (1) Antiferromagnetic order
develops near the vortex center, where the d-wave order parameter is suppressed, 
and exhibits interesting structure within the magnetic unit cell.
(2) The length scale for decay of AFM order is larger than the 
superconducting coherence length. (3) Coexistence of dSC + AFM order
induces $\pi$-triplet order in the vortex core region. (4) Competition between dSC and
AFM splits the zero bias peak in the tunneling conductance.
Experimental data strongly suggest that, at the vortex core, the low energy
states near gap nodes are moved to higher energies. We believe that such 
physics can not be fully captured within mean-field theory.
Dynamic AFM fluctuations and strong correlation effects should be included
to make a more detailed comparison to the experimental data.

\bigskip

\noindent
{\bf Appendix}

\medskip

{\bf Self-consistent Parameters at the Boundary}

\smallskip

The boundary terms corresponding to Eq.~(\ref{eq:Kdefn}) will be modified by
making
the following replacement on the left hand side of Eq.~(\ref{eq:Kdefn}),
\begin{eqnarray}
u_{i\pm \hat x,\alpha}^n({\bf k}) \rightarrow u_{i\pm \hat x \mp N_x\hat x,
\alpha}^n({\bf k}) e^{\mp i k_x N_x} e^{\mp i a i_y N_x}\nonumber \\
u_{i\pm \hat y,\alpha}^n({\bf k}) \rightarrow u_{i\pm \hat y \mp N_y\hat y,
\alpha}^n({\bf k}) e^{\mp i k_y N_y} e^{\mp i a i_x}.\nonumber \\
\end{eqnarray}
and the definition of $\Delta_{{\bf q}n}^{\delta}(i)$ will be modified in the
boundary as following:
\begin{eqnarray}
\Delta_{{\bf q}n}^{\pm \hat x}(i) &=& \frac{Je^{\pm i a i_y N_x}}{4{\cal N}}
\left[\left\{
e^{\mp i q_x N_x} u_{i\pm \hat x \mp N_x\hat x,\downarrow}^n({\bf q})v_{i,
\uparrow}^{n*}({\bf q})\right.\right. \nonumber \\
&+&\left.\left.  e^{\pm i k_x N_x} u_{i,\downarrow}^n({\bf q})v_{i\pm \hat x
\mp N_x\hat x,\uparrow}^{n*}({\bf q})
\right\}
(1-f_{{\bf q} n\downarrow}) \right. \nonumber \\
&+&\left. \left\{e^{\mp i k_x N_x} u_{i,\uparrow}^n({\bf q})v_{i\pm \hat x \mp
N_x\hat x,\downarrow}^{n*}({\bf q})\right.\right. \nonumber \\
&+&\left.\left. e^{\pm i k_x N_x} u_{j,\uparrow}^n({\bf q})
v_{i,\downarrow}^{n*}({\bf q}) \right\}
f_{{\bf k} n\uparrow} \right]\nonumber \\
\Delta_{{\bf q}n}^{\pm \hat y}(i) &=& \frac{J}{4{\cal N}}
\left[\left\{
e^{\mp i q_y N_y} u_{i\pm \hat x \mp N_y\hat y,\downarrow}^n({\bf q})v_{i,
\uparrow}^{n*}({\bf q})\right.\right. \nonumber \\
&+&\left.\left.  e^{\pm i q_y N_y} u_{i,\downarrow}^n({\bf q})v_{i\pm \hat x
\mp N_y\hat y,\uparrow}^{n*}({\bf q})
\right\}
(1-f_{{\bf q} n\downarrow}) \right. \nonumber \\
&+&\left. \left\{e^{\mp i q_y N_y} u_{i,\uparrow}^n({\bf q})v_{i\pm \hat x \mp
N_y\hat y,\downarrow}^{n*}({\bf q})\right.\right. \nonumber \\
&+&\left.\left. e^{\pm i q_y N_y} u_{j,\uparrow}^n({\bf q})
v_{i,\downarrow}^{n*}({\bf k}) \right\}
f_{{\bf q} n\uparrow} \right]\nonumber \\
\end{eqnarray}
Self-consistency conditions for Hartree and Fock shifts (equivalent for
Eq.~(\ref{eq:SCdm})) are not given here; they are easily derivable in our
formalism.

\vskip 0.3cm

{\bf Acknowledgements}: We would like to thank Eugene Kim, 
Henrik Bruus and, in particular, Shoucheng Zhang for many helpful discussions
and suggestions.  This work was supported by NSERC, the Canadian Institute
for Advanced Research and SHARCNet. 



\begin{references}

\bibitem{caroli} C. Caroli, P. G. de Gennes and J. Matricon, Phys. Lett.
{\bf 9}, 307 (1964).
\bibitem{gygi} F. Gygi and M. Schluter, Phys. Rev. {\bf B 43}, 7609 (1991).
\bibitem{hess} H. Hess, R. B. Robinson and J. V. Waszczak, Phys. Rev. Lett.,
{\bf 64}, 2711 (1990).
\bibitem{wang}  Y. Wang and A. H. MacDonald, Phys. Rev. {\bf B 52}, R3876
(1995).
\bibitem{ichioka} M. Ichioka, N. Hayashi, N. Enomoto and K. Machida,
Phys. Rev. {\bf B 53}, 15316 (1996).
\bibitem{maggio} I. Maggio-Aprile, Ch. Renner, A. Erb, E. Walker and
\O. Fischer, Phys. Rev. Lett., {\bf 75}, 2754 (1995).
\bibitem{pan}  S.H. Pan {\it et al.}, Phys. Rev. Lett., {\bf 85}, 1536 (2000).
\bibitem{franz} M. Franz and Z. Tesanovi\'{c}, Phys. Rev. {\bf B 63},
64516 (2001).
\bibitem{han2} Q.-H. Wang, H. H. Han and D.-H. Lee, Phys. Rev. Lett., {\bf 87},
167004 (2001).
\bibitem{han} J. H. Han and D.-H. Lee, Phys. Rev. Lett., {\bf 85}, 1100 (2000).
\bibitem{franz2} M. Franz and Z. Tesanovi\'{c}, Phys. Rev. Lett., {\bf 80},
4763 (1998).
\bibitem{wu} C. Wu, T. Xiang and Z.-B. Su, Phys. Rev. {\bf B 62}, 14427 (2000).
\bibitem{berthod} C. Berthod and B. Giovannini, Phys. Rev. Lett., {\bf 87},
277002 (2001).
\bibitem{arovas} D. P. Arovas, A. J. Berlinsky, C. Kallin and S.-C. Zhang,
Phys. Rev. Lett., {\bf 79}, 2871 (1997). 
\bibitem{andersen} B. M. Andersen, H. Bruus and P. Hedegaard, Phys. Rev.
{\bf B61}, 6298 (2000).
\bibitem{sachdev} E. Demler, S. Sachdev and Y. Zhang, Phys. Rev. Lett.,
{\bf 87}, 67202 (2001); S. Sachdev and S.-C. Zhang, Science, {\bf 295},
452 (2002); Y. Zhang, E. Demler and S. Sachdev, arXiv:cond-mat/0112343;
A. Polkovnikov, M. Vojta and S. Sachdev, arXiv:cond-mat/0203176.
\bibitem{ting1} J.-X. Zhu and C. S. Ting, Phys. Rev. Lett., {\bf 87}, 147002
(2001).
\bibitem{lake} B. Lake {\it et al.}, Science, {\bf 291}, 1759 (2001).
\bibitem{hoffman} J. E. Hoffman {\it et al.}, Science, {\bf 295}, 466 (2002).
\bibitem{khayk} B. Khaykovich {\it et al.}, arXiv:cond-mat/0112505.
\bibitem{mitro} V. F. Mitrovic {\it et al.}, arXiv:cond-mat/0202368;
 V. F. Mitrovic {\it et. al.}, Nature, {\bf 413}, 501, (2001).
\bibitem{miller} R. I. Miller {\it et al.}, Phys. Rev. Lett., {\bf 88}, 137002
(2002).
\bibitem{zhu} J.-X. Zhu, Ivar Martin and A. R. Bishop, arXiv:cond-mat/0201519.
\bibitem{franz3} M. Franz, D. E. Sheehy and Z. Teanovi\'{c}, Phys. Rev. Lett.,
{\bf 88}, 257005, (2002).
\bibitem{zhang2} Han-Dong Chen, Jiang-Ping Hu, Sylvain Capponi, Enrico Arrigoni
and Shou-Cheng Zhang, arXiv:cond-mat/0203332.
\bibitem{zhang} S.-C. Zhang, Science, {\bf 275}, 1089 (1997).
\bibitem{tremblay} B. Kyung and A. M. S. Tremblay, arXiv:cond-mat/0204500.
\bibitem{zhangcomm} S.-C. Zhang, constraint reference.
\bibitem{ting2} Y. Chen and C. S. Ting, arXiv:cond-mat/0112369.

\bibitem{john}
S. Alama, A. J. Berlinsky, L. Bronsard and T. Giorgi,
Phys. Rev. {\bf B 60}, 6901 (1999).

\bibitem{huzhang}
J-P. Hu and S-C. Zhang, arXiv:cond-mat/0108273.

\bibitem{bamsoo}
B. Kyung, Phys. Rev. {\bf B 62}, 9083 (2000).

\bibitem{andreev}
A. F. Andreev, Zh. Eksp. Teor. Fiz. {\bf 46}, 182 (1964); G.E. Blonder,
M. Tinkham and T. M. Klapwijk, {\bf B 25}, 4515 (1982).

\bibitem{murakami}
M. Murakami and H. Fukuyama, J. Phys. Soc. Jpn. {\bf 67}, 2784 (1998).

\bibitem{matsumoto}
M. Matsumoto and H. Shiba, J. Phys. Soc. Jpn. {\bf 64}, 1703 (1995).

\bibitem{matsumoto2}
M. Matsumoto and H. Shiba, J. Phys. Soc. Jpn. {\bf 64}, 3384 (1995);
{\bf 64}, 4867 (1995).

\bibitem{sig}
M. Sigrist, D. B. Bailey and R. B. Laughlin, Phys. Rev. Lett., {\bf 74}, 3249 (1995).

\bibitem{rainer}
D. Rainer, H. Burkhardt, M. Fogelstrom and J. A. Sauls, arXiv:cond-mat/9712234.

\bibitem{buchh}
L. J. Buchholtz, M. Palumbo, D. Rainer and J. A. Sauls,
J. Low. Temp. Phys. {\bf 101}, 1079 (1995); {\bf 101}, 1099 (1995).

\end{references}
\end{document}